\documentstyle[aps,prl,twocolumn,psfig]{revtex}
\begin{document}

\title{Critical exponents of the quantum phase transition in a planar
antiferromagnet}
\author{Matthias Troyer, Masatoshi Imada and Kazuo Ueda}
\address{
Institute for Solid State Physics, University of Tokyo, Roppongi
7-22-1, Tokyo 106, Japan }

\maketitle

\begin{abstract}
We have performed a large scale quantum Monte Carlo study of the
quantum phase transition in a planar spin-1/2 Heisenberg
antiferromagnet with CaV$_4$O$_9$ structure. We obtain a dynamical
exponent $z=1.018\pm0.02$. The critical exponents $\beta$, $\nu$ and
$\eta$ agree within our errors with the classical 3D $O(3)$ exponents,
expected from a mapping to the nonlinear sigma model. This confirms
the conjecture of Chubukov, Sachdev and Ye [Phys. Rev. B {\bf 49},
11919 (1994)] that the Berry phase terms in the planar Heisenberg
antiferromagnet are dangerously irrelevant.
\end{abstract}
 
\vskip 5mm
Instead of classical transitions controlled by temperature $T$ a
quantum phase transition between a symmetry broken phase with
long-range N\`eel order and a quantum disordered state with
a finite spin excitation gap may be realized at $T=0$ by controlling a
parameter $g$ to increase quantum fluctuations. Criticalities around
such quantum phase transitions at $g=g_c$ may reflect inherent quantum
dynamics of the system and yield unusual universality classes with
rich physical phenomena.

The most prominent example are the high temperature
superconductors. There the quantum spin fluctuations are thought to
lead to $d$-wave superconductivity as soon as antiferromagnetism is
suppressed by hole doping.

In this letter we want to discuss a two dimensional quantum Heisenberg
antiferromagnet (2D QAFM) that exhibits such a quantum phase
transition. We will present strong numerical support for the
conjecture that the 2DQAFM is in the same universality class as the
quantum nonlinear sigma model (QNL$\sigma$M) even when Berry phase
terms are present. 

The universality class is characterized by the
critical exponents. Approaching the quantum critical point from the
disordered side the spatial correlation length diverges with the
correlation length exponent $\nu$.  The space and time dimensions are
however not necessarily equivalent, and the correlation length in the
time direction diverges in general with a different exponent $z\nu$,
where $z$ is the dynamical exponent. In a Lorentz invariant system
space and time directions are equivalent and $z=1$.  Related to the
divergence of the correlation length is a vanishing of the spin
excitation gap with the same exponent $z\nu$.  When passing through
the critical point long range order is established.  The order
parameter in the case of a N\'eel ordered antiferromagnet is the
staggered magnetization $m_s$.  Near the critical point $m_s$ vanishes
with the order parameter exponent $\beta$.  At the critical point
itself the real space staggered spin correlation shows a power-law
falloff with power $2-d-z-\eta$, where $\eta$ is the correlation
exponent. These three exponents are related by the usual scaling law
\begin{equation}
\label{eq:scaling}
2\beta = (d+z-2+\eta)\nu,
\end{equation}
where the effective dimension is $d+z$ in a quantum system.

Quantum critical behavior of a planar antiferromagnet has been
intensively studied by a number of groups. Most analytic calculations
are based on the QNL$\sigma$M. The critical exponents of the
QNL$\sigma$M can be determined from simple symmetry, universality and
scaling arguments \cite{CHN,CSY}.  Lorentz invariance implies that
$z=1$.  Furthermore the 2D QNL$\sigma$M is equivalent to the 3D
classical sigma model.  This in turn is in the universality class of
the 3D classical O(3) model, or the classical 3D Heisenberg
ferromagnet. The exponents $\beta$, $\nu$ and $\eta$ should thus be
the same as the well known classical exponents of these models (see
Tab. I).

Chakravarty, Halperin and Nelson have discussed the phase
diagram of a planar Heisenberg antiferromagnet by using the
QNL$\sigma$M.  In their discussions they concentrate on the ordered
phase and describe it as a classical 2D antiferromagnet with
renormalized parameters.

Chubukov, Sachdev and Ye have investigated the quantum critical regime
of the QNL$\sigma$M in close detail.  They make some further
predictions based on scaling arguments. On the ordered side the spin
stiffness $\rho_s$ vanishes as
\begin{equation}
\rho_s\propto(g_c-g)^{(d+z-2)\nu}= (g_c-g)^{\nu},
\end{equation}
where the second equivalence comes from the prediction that $z=1$.
They also predict that the uniform susceptibility at the critical point
is universal:
\begin{equation}
\chi_u = \Omega_1(\infty)\left({g\mu_B\over \hbar c}\right)^2 T.
\end{equation}
Here $c$ is the spin wave velocity and
$\Omega_1(\infty)$ a universal constant. Estimates for $\Omega_1(\infty)$
are listed in Tab. II.

The spin wave velocity $c$ scales as
\begin{equation}
c\propto(g-g_c)^{\nu(z-1)}
\end{equation}
and is thus regular at the critical point if $z=1$.

The equivalence of the 2D QAFM to the 2D QNL$\sigma$M however is still
an open question because of the existence of Berry phase terms in the
QAFM that are not present in the QNL$\sigma$M \cite{haldane}.  It has
been argued that these terms cancel in special cases, such as in the
bilayer model \cite{sandvik,Duin}.  Then it is plausible that the
quantum phase transition is in the same universality class as the
QNL$\sigma$M. This was confirmed by quantum Monte Carlo calculations
of Sandvik and coworkers
\cite{sandvik2,sandvik,sandvik3}. They have investigated the finite
size scaling of the ground state structure factor and susceptibilities
on lattices with up to $10 \times 10\times 2$ spins. Although these
lattices are quite small they still found good agreement of the
exponents $z$ and $\eta$ with the QNL$\sigma$M predictions
\cite{sandvik2,sandvik} (see Tab. I). In another study Sandvik {\it et
al.} \cite{sandvik3} have investigated finite temperature properties
of the bilayer QAFM on larger lattices and also found good agreement
with the QNL$\sigma$M predictions. In the absence of Berry phase terms
the equivalence of the QAFM and the QNL$\sigma$M is quite well
established by these simulations.

But in general these Berry phase terms exist. Chakravarty {\it et al.}
argue that they can change the critical behavior and lead to different
exponents\cite{CHN,Chakravarty}.  Chubukov {\it et al.} on the other
hand argue that the Berry phase terms are dangerously
irrelevant\cite{CSY} and do not influence the critical
behavior. Previous numerical simulations on dimerized square lattices
\cite{sandvik2,katoh} are indeed not consistent with the QNL$\sigma$M
predictions. The reliability of the results however is questionable
because of the restriction to very small lattices of $12\times 12$
spins and because of complications with scaling arising from
inequivalent spatial directions.  On the other hand the discrepancy
could be an effect of the Berry phase terms that are present in the
dimerized square lattice but probably not in the bilayer.

Using the new quantum cluster algorithms \cite{evertz,beard} we could
simulate much larger lattices at lower temperatures. On these larger
lattices we find perfect agreement with predictions made based upon
the 2D QNL$\sigma$M despite the presence of Berry phase terms.

As the universality class of a phase transition does not depend on the
microscopic details of the lattice structure we are free to choose the
best lattice for our purposes. We have chosen the ${\rm CaV}_4{\rm
O}_9$ lattice, a 1/5-th depleted square lattice depicted in Fig. 1 for
our calculations. There are three reasons for this choice. Firstly the
Berry phase terms are present on this lattice \cite{Sachdev}. Next
both space directions are equivalent, in contrast to the dimerized
square lattice \cite{sandvik2,katoh}. This makes the scaling analysis
easier. Finally at the quantum critical point all the couplings are
nearly equal in magnitude, which is also optimal from a numerical
point of view. We have performed our simulations on square lattices
with $N=8n^2$ spins, where $n$ is an integer. Our largest lattices
contained 20 000 spins. For the following discussion it is useful to
introduce the linear system size $L$ in units of the bond lengths $a$
of the original square lattice: $L\equiv \sqrt{5N/4} a$.

The phase diagram of this lattice has been discussed in detail in Ref.
\cite{cvo}.  By removing every fifth spin we obtain a lattice 
consisting of four-spin plaquettes linked by dimer bonds. We label the
couplings in a plaquette $J_0$ and the inter-plaquette couplings
$J_1$.  By controlling the ratio of these couplings $J_1/J_0$ we can
tune from N\'eel order at $J_1=J_0$ to a quantum disordered
``plaquette RVB'' ground state with a spin gap $\Delta=J_0$ at
$J_1=0$.

At some intermediate coupling ratio $(J_1/J_0)_c$ the systems has a
quantum phase transition. The first step in the determination of the
critical behavior is a high precision estimate of the critical point
$g_c$. We have calculated the second moment correlation length $\xi_L$
on systems of various sizes $L$. The temperature was chosen to be $k_B
T=J_0 a/L$, keeping the finite $2+1$ dimensional system in the cubic
regime. From standard finite size scaling arguments it follows that
this correlation length $\xi_L$ scales proportional to the system size
$L$ at criticality. We have calculated the ratio $\xi_L/L$ (shown in
Fig. 2) for a variety of couplings and system sizes up to $N = 9600$
and have determined the critical coupling to be
$(J_1/J_0)_c=0.939\pm0.001$.

Next we have calculated the finite size scaling of both the staggered
structure factor $S({\bf Q}) = L^2m_s$ and of the corresponding
staggered susceptibility. At criticality they scale like
\begin{eqnarray}
S({\bf Q})& \propto& L^{2-z-\eta} \label{eq:fss1}\\
\chi_{s}& \propto& L^{2-\eta} \label{eq:fss2}
\end{eqnarray}
The temperature was chosen to be $k_B T=J_0 a/(4L)$. This was low
enough to see the ground state properties on the finite lattice.  By
fitting our results shown in Fig. 3 we obtain the estimates
$z=1.018\pm0.02$ and $\eta=0.015 \pm 0.020$. This is perfectly
consistent with the Lorentz invariance ($z=1$) expected from a mapping
to the QNL$\sigma$M. We will discuss $\eta$ below together with the
other exponents. From these fits it is also obvious that at least
$N=800$ spins are necessary to obtain good scaling.

The remaining exponents $\beta$ and $\nu$ are best calculated from the
magnetization $m_s$ and the spin stiffness $\rho_s$ on the ordered
side. Good estimates for $m_s$ and $\rho_s$ can be obtained from the
Hasenfratz-Niedermayer equations \cite{HN}.  These authors have
calculated the {\it exact} finite-size and finite-temperature values
of the low-temperature uniform and staggered susceptibilities $\chi_u$
and $\chi_s$ for the ordered phase of a 2D QAFM on a lattice with the
symmetries of a square lattice.  Their equations, determined by chiral
perturbation theory, are correct for the low temperature regime $k_B T
\ll 2\pi\rho_s$ with cubic geometry $k_B TL/ \hbar c \approx 1$.  Up
to second order in $T$ (or $1/L$ respectively) the susceptibilities
are universal, determined by only three parameters: the staggered
magnetization $m_s$, the spin stiffness $\rho_s$ and the spin wave
velocity $c$. Two high precision auantum Monte Carlo studies have
confirmed their equations for the square lattice QAFM
\cite{beard,wiese}.

We have calculated the susceptibilities for a wide range of couplings
$0.95 < J_1/J_0 < 1.1$, lattice sizes $800 < N < 16200$ and temperatures
$0.006 < T/J_0 < 0.1$. The fits to the Hasenfratz-Niedermayer
equations are all excellent, with $\chi^2/\mbox{\rm d.o.f.}\approx
1.5$.  This is another confirmation of the universality of the
Hasenfratz-Niedermayer equations.  From the fits we obtain the
staggered magnetization $m_s$, the spin stiffness $\rho_s$ and the
spin wave velocity $c$. The exponents $\beta$ and $\nu$ can then
be obtained in a straightforward way (see Fig. 4) and are listed in
Tab. I.

Let us now discuss the results. First we observe that the exponents
satisfy the scaling relation Eq. (\ref{eq:scaling}), confirming the
validity of the scaling ansatz for this quantum phase transition. The
exponents $\beta$, $\nu$ and $\eta$ are in excellent agreement with
the exponents of the 3D classical O(3) or Heisenberg model. They are
however incompatible with the mean field exponents suggested by Katoh
and Imada from their calculations on small lattices.

Assuming Lorentz invariance ($z=1$) we can improve our estimates for
the other exponents. The agreement of the improved estimates with the
3D O(3) exponents becomes even better. We can rule out not only the
mean field universality class suggested by Ref. \cite{katoh}, but also
the Ising universality class (see Table I for a comparison).

This excellent agreement is a strong numerical support for the
conjecture of Chubukov, Sachdev an Ye \cite{CSY} thet the Berry phase
terms in the 2D QAFM are indeed dangerously irrelevant. To further
confirm their predictions we have calculated the uniform
susceptibility close to criticality down to $T=0.02$, more than an
order of magnitude lower than Ref. \cite{sandvik}. We have
extrapolated the finite size results on lattices with up to $N=20 000$
spins to the thermodynamic limit. Looking for the coupling at which a
linear behavior occurs gives an independent estimate of the critical
point: $(J_1/J_0)_c=0.939\pm0.002$, in excellent agreement with the
above estimate. The linear slope is $\Omega_1(\infty)(J_0/\hbar c)^2
=0.238
\pm 0.003 $.  By extrapolating the spin wave velocity determined in
the ordered phase by the Hasenfratz-Niedermayer fit to the critical
point we get $\hbar c/J=1.04\pm0.02$ and thus $\Omega_1(\infty) =
0.26\pm0.01$, again in excellent agreement with Chubukov {\it et al.}
\cite{CSY} (see Tab. II).

To summarize, we have calculated the critical exponents of the quantum
critical point in a planar antiferromagnet by a large scale quantum
Monte Carlo study.  Our exponents agree perfectly with predictions
made by a mapping to the 2D quantum nonlinear sigma model. The
dynamical exponent is $z=1.018\pm0.02$, consistent with Lorentz
invariance. The other exponents agree with the 3D classical O(3)
exponents. The conjecture that all quantum phase transitions between a
N\'eel ordered and a quantum disordered state in 2D Heisenberg
antiferromagnets, whether they contain Berry phase terms or not, are
in the same universality class as the quantum nonlinear sigma model is
strongly supported.

We want to thank the Computer Center of the university of Tokyo for
giving us the ability to use their 1024-node massively parallel
Hitachi SR 2201 supercomputer. Being able to use this fast computer
has enabled us to perform the simulations reported here.  We are
grateful to J.-K. Kim, D.P. Landau, S. Sachdev, A.W. Sandvik and
U.-J. Wiese for interesting discussions.


\begin{figure}
\begin{center}
\mbox{\psfig{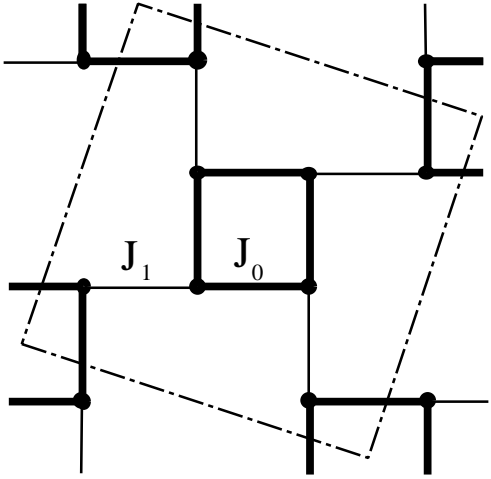}}
\end{center}
\caption{Lattice structure of the 1/5-th depleted square lattice
of CaV$_4$O$_9$. The dashed square indicates the eight spin unit cell
used in our calculations.}
\end{figure}

\begin{figure}
\begin{center}
\mbox{\psfig{figure=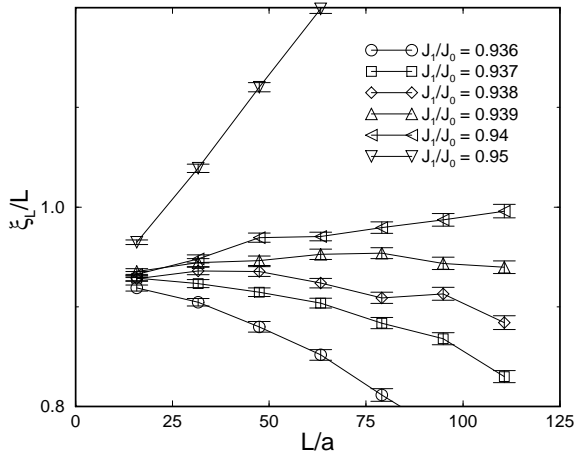,width=\hsize}}
\end{center}
\caption{Plot of the ratio of correlation length divided by system size
$\xi_L/L$. At the critical point the correlation length calculated in a
finite system is proportional to the system size. This is the case for
$g=J_1/J_0=0.939\pm0.001$.}
\end{figure}

\begin{figure}
\begin{center}
\mbox{\psfig{figure=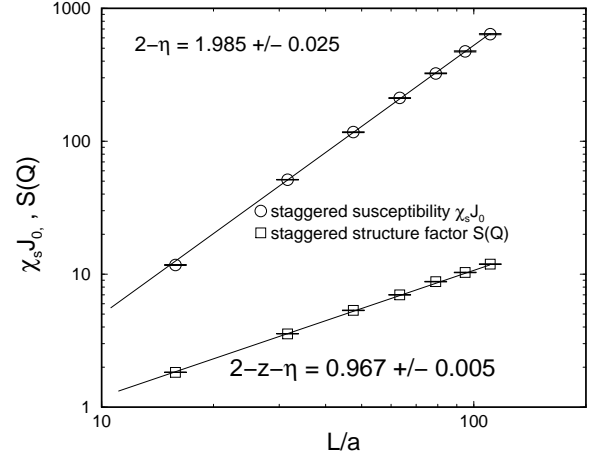,width=\hsize}}
\end{center}
\caption{Finite size scaling of the staggered structure factor and
susceptibility at the critical point. The straight lines are fits to
the finite size scaling forms Eqns. (\protect{\ref{eq:fss1}}) and
(\protect{\ref{eq:fss2}}).}
\end{figure}

\begin{figure}
\begin{center}
\mbox{\psfig{figure=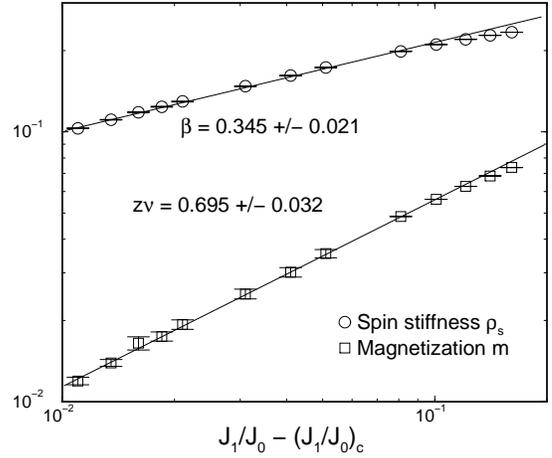,width=\hsize}}
\end{center}
\caption{Staggered magnetization $m_s$ and spin stiffness $\rho_s$
calculated by a fit of the low temperature susceptibilities on finite
lattices to the Hasenfratz-Niedermayer equations
\protect{\cite{HN}}. The straight lines are fits used to obtain the
exponents $\beta$ and $\nu$.}
\end{figure}

\eject
\widetext
\begin{table}
\caption{Critical exponents $\beta$, $\nu$, $\eta$ and $z$.
Listed are both the estimates without making any assumption for $z$,
and the best estimate if Lorentz invariance ($z=1$) is assumed.  For
comparison the exponents of the 3D classical Heisenberg (O(3)) model,
the 3D Ising model and the 2D quantum mean field exponents are listed.
The errors given include the uncertainties in the critical point.}
\begin{tabular}{l|cccc}
model & $\nu$ & $\beta$ & $\eta$ & $z$ \\
\hline
2D QAFM & $ 0.685 \pm 0.035 $ & $ 0.345 \pm 0.025 $ & $ 0.015 \pm
0.020 $ & $ 1.018 \pm 0.02 $\\ Lorentz invariant 2D QAFM & $ 0.695 \pm
0.030 $ & $ 0.345 \pm 0.025 $ & $ 0.033 \pm 0.005 $ & 1 (assumption)\\
\hline
bilayer QAFM \protect{\cite{sandvik2}} & & & $0.03 \pm 0.01$ &$ 1.08
\pm 0.05$ \\
\hline
3D O(3) \protect{\cite{expO3}} & $ 0.7048 \pm 0.0030 $ & $ 0.3639 \pm
0.0035 $ & $ 0.034 \pm 0.005 $ & --- \\ 3D Ising
\protect{\cite{expising}} & $ 0.6294 \pm 0.0002 $ & $ 0.326 \pm 0.004
$ & $ 0.0327 \pm 0.003 $ & --- \\ mean field & 1 & 1/2 & 0 & 1 \\
\end{tabular}
\end{table}
\narrowtext

\begin{table}
\caption{Universal prefactor $\Omega_1(\infty)$ in the linear
temperature dependence of the uniform susceptibility at
criticality. Listed are the results for the quantum nonlinear sigma
model in a $1/N$ expansion, the results by classical Monte Carlo
simulation on a 3D classical rotor model and the result of the present
study.}

\begin{tabular}{ll|c}
method & Ref. & $\Omega_1(\infty)$ \\
\hline
1/N expansion & \protect{\cite{CSY}} & 0.2718 \\
classical Monte Carlo & \protect{\cite{CSY}} & $0.25\pm0.04$ \\
quantum Monte Carlo & this study & $0.26\pm0.01$\\
\end{tabular}
\end{table}
\end{document}